\documentclass[doublecol]{epl2}
 \usepackage{etex} \usepackage[english]{babel} \usepackage[T1]{fontenc} \usepackage[utf8]{inputenc}
  \usepackage[babel]{csquotes}
\usepackage{amsmath} \usepackage{graphicx}
 
 \newcommand{\BV}{Brunt-V\"ais\"al\"a\ }    
 \newcommand{\be}{\begin{equation}}         
 \newcommand{\ee}{\end{equation}}         
\definecolor{orange}{cmyk}{0,0.5,1,0}

\title{Connecting Large-Scale Velocity and Temperature Bursts with Small-Scale Intermittency in Stratified Turbulence}
\shorttitle{Connecting Large-Scale Velocity and Temperature Bursts with Small-Scale Intermittency in Stratified Turbulence}

\author{F. Feraco $^{1,2}$, R. Marino $^{1}$, L. Primavera $^{2}$, A. Pumir $^{3}$, P.D. Mininni $^{4}$, D. Rosenberg$^{5}$, A. Pouquet$^{6}$, R. Foldes$^{7}$, E. L\'ev\^eque $^{1}$, E. Camporeale$^{8}$, S. Cerri$^{9}$  H. Charuvil Asokan$^{9}$, J.L. Chau$^{9}$, J.P. Bertoglio $^{1}$, P. Salizzoni$^{1}$, M. Marro$^{1}$}
\shortauthor{F. Feraco, R. Marino, L. Primavera, A. Pumir, P.D. Mininni {\it et al.}}

\institute{%
\inst{1} \quad Laboratoire de M\'ecanique des Fluides et d'Acoustique, CNRS, \'Ecole Centrale de Lyon,
         Universit\'e Claude Bernard Lyon~1, INSA de Lyon, F-69134 \'Ecully, France\\
\inst{2}\quad Dipartimento di Fisica, Universit\`a della Calabria, Italy\\
\inst{3} \quad \'Ecole Normale Sup\'erieure de Lyon, Lyon France\\
\inst{4} \quad Departamento de F\'{\i}sica, Facultad de Ciencias Exactas y Naturales, 
        Universidad de Buenos Aires, and IFIBA, CONICET, Buenos Aires 1428, Argentina\\
\inst{5} \quad 288 Harper St., Louisville, CO 80027, USA \\
\inst{6} \quad Laboratory for Atmospheric and Space Physics, University of Colorado, Boulder, CO 80309, USA, and National Center for Atmospheric Research, P.O.~Box 3000, Boulder, CO 80307, USA\\
\inst{7}\quad Universit\`a degli studi dell'Aquila, Italy\\
\inst{8} \quad CIRES, University of Colorado, Boulder, CO 80309, USA, and
Centrum Wiskunde \& Informatica, 1098 XG Amsterdam, The Netherlands\\
\inst{9} \quad Department of Astrophysical Sciences, Princeton University, 4 Ivy Lane, Princeton, NJ 08544, USA\\
\inst{10} \quad Leibniz Institute of Atmospheric Physics, University of Rostock, K\"uhlungsborn, Germany\\}

\abstract{ 
Non-Gaussian statistics of large-scale fields 
are routinely observed in data from atmospheric and oceanic  
campaigns and global models. Recent direct numerical simulations (DNSs) showed that large-scale intermittency in stably stratified flows is due to the emergence of sporadic, extreme events in the form of bursts in the vertical velocity and the temperature. This phenomenon results from the interplay between waves and turbulent motions, affecting mixing. We provide evidence of the enhancement of the classical small-scale (or internal) intermittency due to the emergence of large-scale drafts, connecting large- and small-scale bursts.
To this aim we analyze a large set of DNSs of the stably stratified Boussinesq equations over a wide range of values of the Froude number ($Fr\approx 0.01-1$).
The variation of the buoyancy field kurtosis with $Fr$ is similar to (though with smaller values than) the kurtosis of the vertical velocity, both showing a non-monotonic trend. 
We present a mechanism for the generation of extreme vertical drafts and vorticity enhancements which follows from the exact equations for field gradients.
}
\pacs{47.55.Hd}{Stratified flows}
\pacs{47.35.Bb}{Gravity waves}
\pacs{42.68.Bz} {Atmospheric turbulence effects}
\begin{document}
\maketitle
\section{Introduction}
\noindent A characteristic 
feature of turbulent flows, including geophysical flows, is the so-called internal (or small-scale) intermittency, producing localized intense variations of the energy dissipation and of field gradients \cite{frisch_80}, as observed in many instances in the  atmosphere \cite{fritts_13}, the ocean \cite{vanharen_15,pearson_18}, and in laboratory experiments and in direct numerical simulations (DNSs). 
The Kolmogorov refined theory~\cite{kolmogorov_62} of homogeneous isotropic turbulence (HIT) relates this phenomenon to the anomalous scalings of the structure functions~\cite{kolmogorov_62}.
Internal intermittency is also at the origin of the non-Gaussian behavior of the probability distribution functions (PDFs) of small-scale turbulent velocity fluctuations. But an accurate modeling of intermittency in turbulence remains a challenging question. Numerous attempts have been made to model this departure from Gaussianity, as for instance with log-normal or log-L\'evy models \cite{she_94}.
When shear waves are present, it was shown that intermittency is again a major feature of the small-scale behavior in these flows \cite{toschi_00}, although the anomalous exponents differ from those obtained in HIT, showing a lack of dependence with the imposed shear.

More generally, intermittent events at large scales have  been observed in clear air turbulence with patches that can span up to 100 km horizontally and 1 km vertically \cite{Bramberger_2020}, in the vertical velocity and temperature in atmospheric mesoscales \cite{lyu_2018}, in the mesosphere-lower thermosphere \cite{chau_17}, in the ocean (in observations \cite{dasaro_07} and models \cite{capet_08}), and in DNSs \cite{rorai_14, feraco_18, sujovolsky_19} where large-scale bursts develop in a certain parameter space. Indeed, \cite{feraco_18} (F18 hereafter) demonstrated that the vertical velocity $w$ and the buoyancy $\theta$ (proportional to potential temperature fluctuations) are highly intermittent at large scales in DNSs of the Boussinesq equations. This phenomenon takes place in a range of the Froude number $Fr$ compatible with values in the atmosphere and the oceans. Large-scale intermittency was found both for Eulerian and Lagrangian velocities, demonstrating the importance of large-scale flow structures. In turn, these structures were shown to be associated with the most unstable regions within the fluid, affecting its mixing properties \cite{feraco_18}. More recently it was shown that large-scale intermittency is present also in rotating stratified turbulent flows, including situations in quasi-geostrophic balance \cite{pouquet_19p, buaria}. 

The purpose of this study is to
connect large- and small-scale intermittency in stably stratified turbulence, using the DNS database of purely stratified flows from F18. 
We analyze fourth-order moments of the Eulerian velocity and buoyancy fields. With these fields we show the connection between intermittent dynamics at different scales, and how intermittency and the emergence of structures in the flow are modulated by the Froude number.

\section{Methods}
\noindent The results presented here are based on several DNSs of the Navier-Stokes equations in the Boussinesq framework in presence of stable stratification. To ensure incompressibility, the velocity field $\mathbf{u}$ satisfies the condition $\nabla \cdot \mathbf{u} = 0$. The dimensionless equations are: 
%
\begin{eqnarray} \label{bequationV} 
\partial_t \textbf u+(\textbf u \cdot\nabla)\textbf u  
&=& - \nabla p - N\theta\hat{z} +  \textbf F +\nu\nabla^2\textbf u \\
\partial_t\theta+\textbf u\cdot\nabla\theta &=& Nw + \kappa\nabla^2\theta,
\label{bequationT} \end{eqnarray}
%
where $\nu$ and $\kappa$ are respectively the kinematic viscosity and the thermal diffusivity. 
For all the runs the Prandlt number is $Pr=\nu/\kappa=1$ with $\nu=10^{-3}$. $N$ is the \BV frequency, associated to the background potential temperature stratification, kept constant throughout the computational domain, thus representing the parameter governing the imposed stable stratification.
The initial conditions consist of vanishing buoyancy fluctuations ($\theta = 0$) and a velocity field with kinetic energy randomly distributed on spherical shells with wavenumbers $k_F \in [2,3]$ in Fourier space. 
A random isotropic mechanical forcing $\textbf F$ is  applied to the velocity field at the same wavenumbers \cite{marino_14}. We define the dimensionless parameters $Re = UL / \nu$ and $\  Fr = U / LN$ respectively as the Reynolds and Froude numbers, where $U$ and $L$ are the flow characteristic velocity and integral scale. 
The buoyancy Reynolds number $R_B \equiv Re \, Fr^2$ measures the relative strength of buoyancy to dissipation and is commonly used to identify regimes where waves ($R_B  \le 10$) or turbulence ($R_B  \ge 10^2$) dominate. 
We integrated the equations numerically using the Geophysical High-Order Suite for Turbulence (GHOST), a hybrid MPI-, OpenMP- and CUDA-parallelized pseudo-spectral code \cite{rosenberg_20} that can generate flows in a triply-periodic domains (as in the runs under study) or with non-periodic boundary conditions in one direction \cite{fontana_20}. 
Seventeen runs were performed on isotropic grids of $512^3$ points, with the size of the periodic three-dimensional computational box equal to $2\pi$, each run with a different value of $N$. The statistics of ${\mathbf u}=(u,v,w)$ and $\theta$ are characterized by their dimensionless fourth-order moment, the kurtosis, \begin{equation}
\label{kurtalfa}
K_{\alpha}=\frac{\langle (\alpha - \bar\alpha)^4 \rangle } { \langle (\alpha - \bar\alpha)^2 \rangle^2} \  , \\  
\end{equation}
with $\alpha=u$, $v$, $w$, $\theta$, or field gradients. Averages were taken in all cases over the entire spatial domain and for $\approx 8$ turnover times $\tau_{NL}=L/U$ after the peak of dissipation was reached. The reference value of the kurtosis for Gaussian processes is 3, thus values $K_{\alpha}>3$ correspond to leptokurtic PDFs with fat tails and a higher probability of extreme values.
As seen in F18, $u$ and $v$ show no large-scale intermittency, their kurtosis never exceeding $3$, and thus with nearly Gaussian PDFs as in HIT, or, in fact, slightly sub-Gaussian up to $Fr\approx 0.2$. However, in F18 it was shown that $w$ and $\theta$ develop strong events in the range $0.07\le Fr \le 0.1$, with  
$K_{w,\theta}>3$. Table~\ref{Tab1} gives 
relevant quantities and the governing parameters for each run.

\begin{table*}[!h] \centering
\setlength\tabcolsep{4.5pt}
\footnotesize
\begin{tabular}{c c c c c c c c c c c c c c c c c c c  p{1cm}}
Run&1&2&3&4&5&6&7&8&9&10&11&12&13&14&15&16&17\\
\hline
\hline
		\rule{0pt}{2.5ex} $Re/10^3$
		&3.9&3.8&3.8&3.8&3.8&3.8&3.9&3.8&3.8&3.8&3.7&3.6&3.0&2.6&2.6&2.8&2.9 \\
		\rule{0pt}{2.5ex} $Fr$
		&0.015&0.026&0.030&0.038&0.044&0.051&0.068&0.072&0.076&0.081&0.098&0.11&0.16&0.19&0.28&0.56&0.93 \\
		\rule{0pt}{2.5ex} $R_B$
		&0.87&2.5&3.4&5.6&7.3&10.2&17.7&19.7&22.1&25.2&35.9&47.5&75.2&90.9&201&895&2560 \\
		\hline\hline
		\rule{0pt}{2.5ex} $K_{u}$ &2.3&2.4&2.3&2.1&2.3&2.3&2.3&2.3&2.3&2.3&2.3&2.3&2.5&2.6&2.9&2.8&2.8 \\
		\rule{0pt}{2.5ex} $K_{v}$ &2.2&2.3&2.2&2.3&2.1&2.0&2.1&2.1&2.1&2.1&2.1&2.1&2.6&2.8&2.7&2.8&2.8 \\
		\rule{0pt}{2.5ex} $K_{w}$ &3.1&3.2&3.1&3.1&3.2&3.6&7.3&8.6&10.4&9.1&8.8&5.3&3.9&3.5&3.3&3.0&2.9 \\
		\rule{0pt}{2.5ex} $K_{\theta}$&3.3&3.4&3.4&3.5&3.5&3.6&4.0&4.3&4.3&4.1&4.1&3.6&3.1&2.9&2.8&2.7&2.7 \\
		\hline\hline
		\rule{0pt}{2.5ex} $K_{\partial_x\theta}$
		&4.7&4.9&6.5&11.7&16.0&45.6&118.0&101.5&112.2&71.7&53.0&28.6&17.5&15.9&15.6&13.6&13.5 \\
		\rule{0pt}{2.5ex} $K_{\partial_y\theta}$
		&5.0&5.2&6.1&14.9&58.7&157.0&165.0&140.0&150.1&88.1&84.2&34.5&18.5&15.3&16.0&13.5&13.1 \\
		\rule{0pt}{2.5ex} $K_{\partial_z\theta}$ &9.8&6.3&6.5&6.4&6.1&6.5&8.7&8.8&10.3&8.7&11.1&8.7&10.2&11.8&16.9&15.3&15.3 \\
		\hline\hline
		\rule{0pt}{2.5ex} $K_{\partial_x u}$ &4.0&4.5&4.0&4.0&7.4&9.9&49.6&37.3&38.6&33.9&26.0&16.3&7.3&6.03&5.5&5.7&5.8 \\ 
	    \rule{0pt}{2.5ex} $K_{\partial_y v}$ &3.9&4.6&4.6&6.1&22.7&59.5&83.5&71.2&57.4&41.2&35.9&19.2&7.5&6.0&5.5&5.7&5.8 \\ 
		\rule{0pt}{2.5ex} $K_{\partial_z w}$ &3.4&3.7&5.3&27.5&67.4&90.1&88.4&73.7&56.7&38.3&25.6&13.0&6.0&5.4&5.2&5.6&5.8 \\ 
        \hline\hline
        \rule{0pt}{2.5ex} $K_{\partial_y u}$ 
        &5.2&5.9&5.8&6.0&7.7&24.3&58.5&55.7&48.2&51.0&44.6&25.1&10.1&8.6&8.1&8.6&8.8 \\ 
	    \rule{0pt}{2.5ex} $K_{\partial_z u}$ 
	    &4.4&3.9&4.0&4.0&4.01&3.9&3.8&3.9&3.9&3.8&4.2&4.3&6.0&6.9&7.5&8.3&8.7 \\ 
		\rule{0pt}{2.5ex} $K_{\partial_x v}$ 
		&4.3&5.0&4.3&5.2&6.6&8.8&60.7&39.4&56.4&41.4&36.5&23.0&10.1&8.2&7.9&8.5&8.7 \\ 
    	\rule{0pt}{2.5ex} $K_{\partial_z v}$ 
	    &3.8&4.2&3.7&3.9&3.8&3.7&3.9&3.8&3.9&3.8&4.0&4.1&5.9&6.9&7.5&8.4&8.6 \\ 
        \rule{0pt}{2.5ex} $K_{\partial_x w}$ 
        &5.0&5.6&8.4&25.1&61.9&176.5&361.6&225.0&258.4&133.8&78.4&35.8&11.3&8.7&7.7&8.3&8.7 \\ 
	    \rule{0pt}{2.5ex} $K_{\partial_y w}$ 
	    &5.4&6.1&10.8&41.4&222.1&354.0&236.0&191.1&199.1&112.1&89.5&36.6&11.6&8.8&7.7&8.5&8.6 \\ 
        \hline\hline
\end{tabular}
\caption{\it Governing parameters for each run, namely the Reynolds number $Re$, the Froude number $Fr$, and the buoyancy Reynolds number $R_B$, and kurtosis of the fields  and their gradients, $K_\alpha$ ($\alpha=u$, $v$, $w$, $\theta$, and spatial derivatives $\partial_x$, $\partial_y$ and $\partial_z$ of all these quantities) for all the runs. Values are averaged for $\approx 8$ turnover times after the peak of dissipation in each run.} 
\label{Tab1}
\end{table*}

\section{Results} \label{S:R}

\begin{figure*}[h!]
 \begin{center}
\hspace*{-0.8cm}
 \includegraphics[width=12cm]{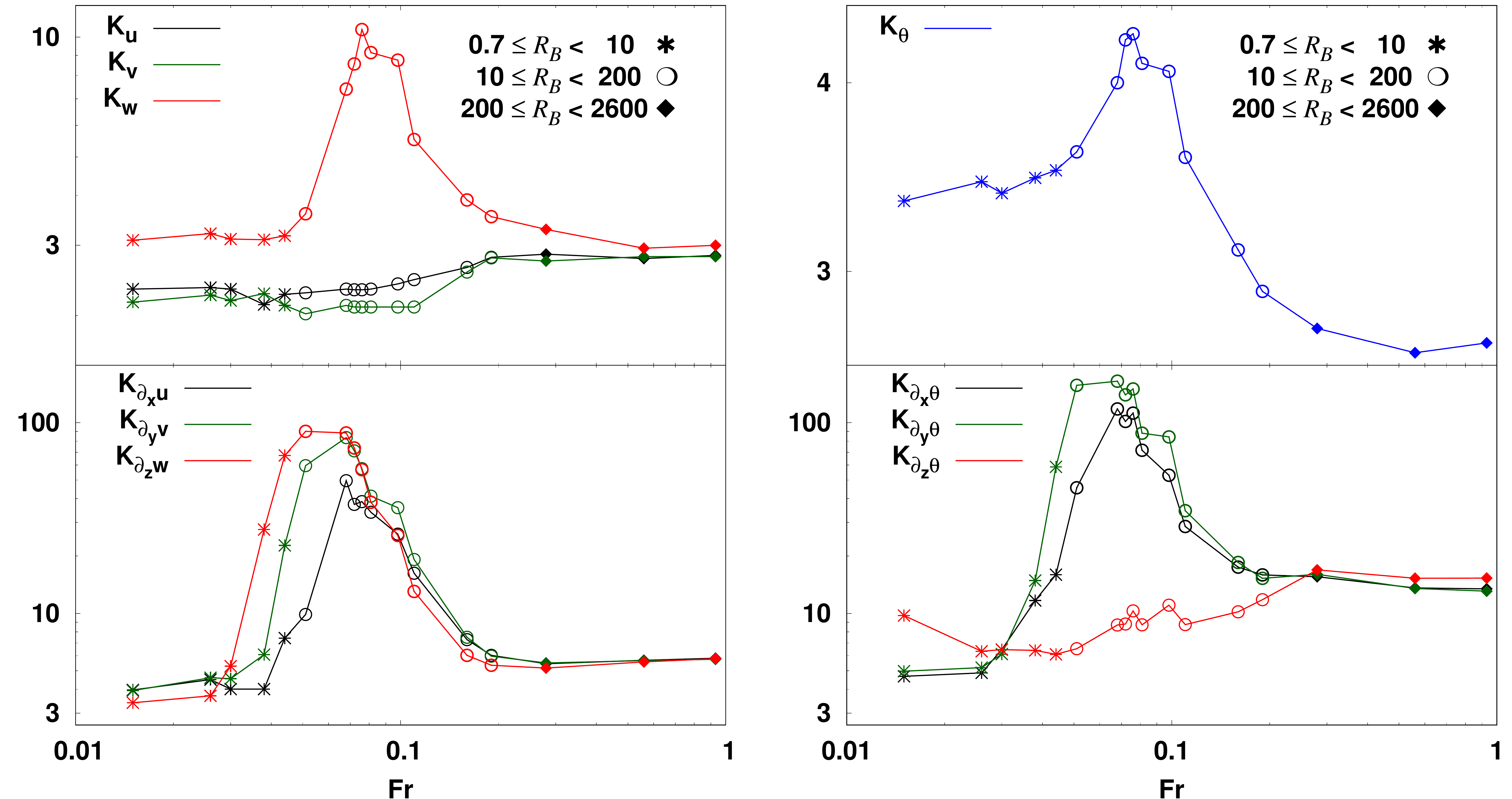}
 \end{center}
 \vskip-0.2truein   
 \caption{\it For all runs in Table~\ref{Tab1}, the figure shows as a function of $Fr$: the kurtosis of $u$, $v$, $w$ (top-left) and $\theta$ (top-right); the kurtosis of the diagonal elements of the velocity gradient tensor $\partial_x u$, $\partial_y v$,  $\partial_z w$ (bottom-left), and of the components of the buoyancy gradient $\partial_x\theta$, $\partial_y\theta$, and $\partial_z\theta$.
 Note that $K_{u}$, $K_{v}$, and $K_{\partial_z\theta}$ do not show any sizable dependence with $Fr$, the first two being always compatible with Gaussian or sub-Gaussian statistics.
 }
 \label{FIG1}
 \vskip-0.15truein      
 \end{figure*}

The Eulerian vertical velocity $w$ kurtosis for runs in Table~\ref{Tab1} follows a non-monotonic dependence on $Fr$ 
with a peak at $Fr\approx 0.08$; see Fig.~\ref{FIG1} (top). The figure also shows that $K_\theta$ exhibits a qualitatively similar dependence with a peak at $Fr \approx 0.08$.  The value of $K_\theta$ at the peak, however, is significantly smaller than that of $K_w$. The buoyancy field $\theta$ is therefore intermittent at large scales, with the emergence of localized bursts making its PDF non-Gaussian in the range $Fr \in [0.07-0.1]$, corresponding to runs $7$ to $12$ in Table~\ref{Tab1}. As a reference, in the atmosphere at horizontal scales of 100 km, $Fr \approx 0.01$ on the average, but values in the range mentioned above are not unusual (e.g., up to $0.5$ in clear air turbulence \cite{Bramberger_2020} with turbulent patches of 1 to 100 km).

\begin{figure} [t!]
 \begin{center} 
 \hspace*{-0.35cm}
 \includegraphics[width=9.1cm]{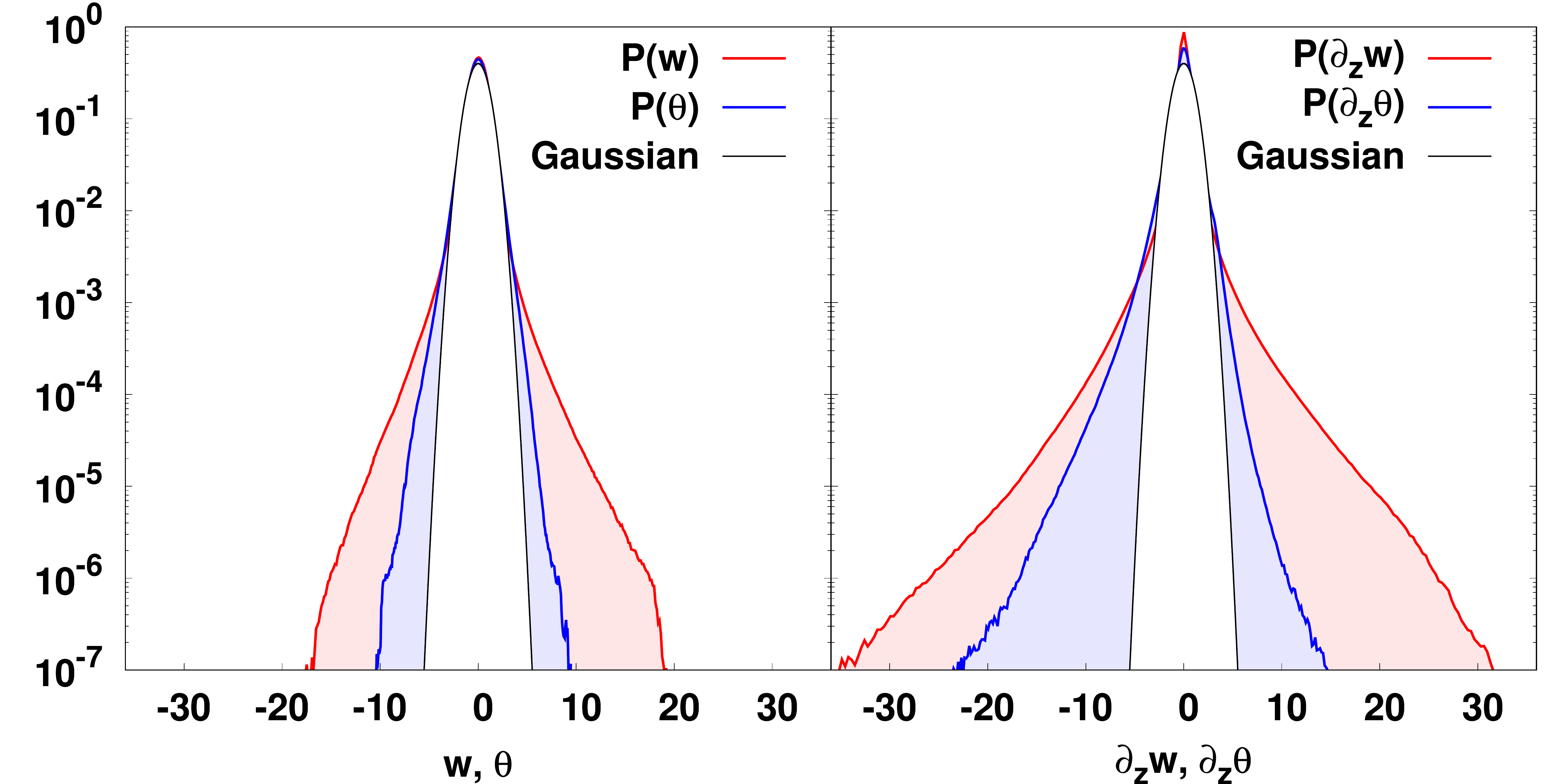}   
 \end{center}
 \vskip-0.2truein   
 \caption{\it (Left) PDFs of $w$ and $\theta$ for run $9$ with $K_w=10.4$, $K_\theta=4.3$. (Right) PDFs of $\partial_z w$ and $\partial_z\theta$ for the same run, with $K_{\partial_z w}=56.7$, $K_{\partial_z\theta}=10.3$. Shaded areas exceeding the Gaussian PDF (black lines) indicate the probability for extreme values in the fields and their gradients. Fields are normalized by their standard deviation to allow direct comparison.
}
 \label{FIG2}
 \vskip-0.15truein       
 \end{figure}

\begin{figure} [t!]
		\begin{center}
			\hspace*{-0.3cm}
			\includegraphics[width=8.8cm]{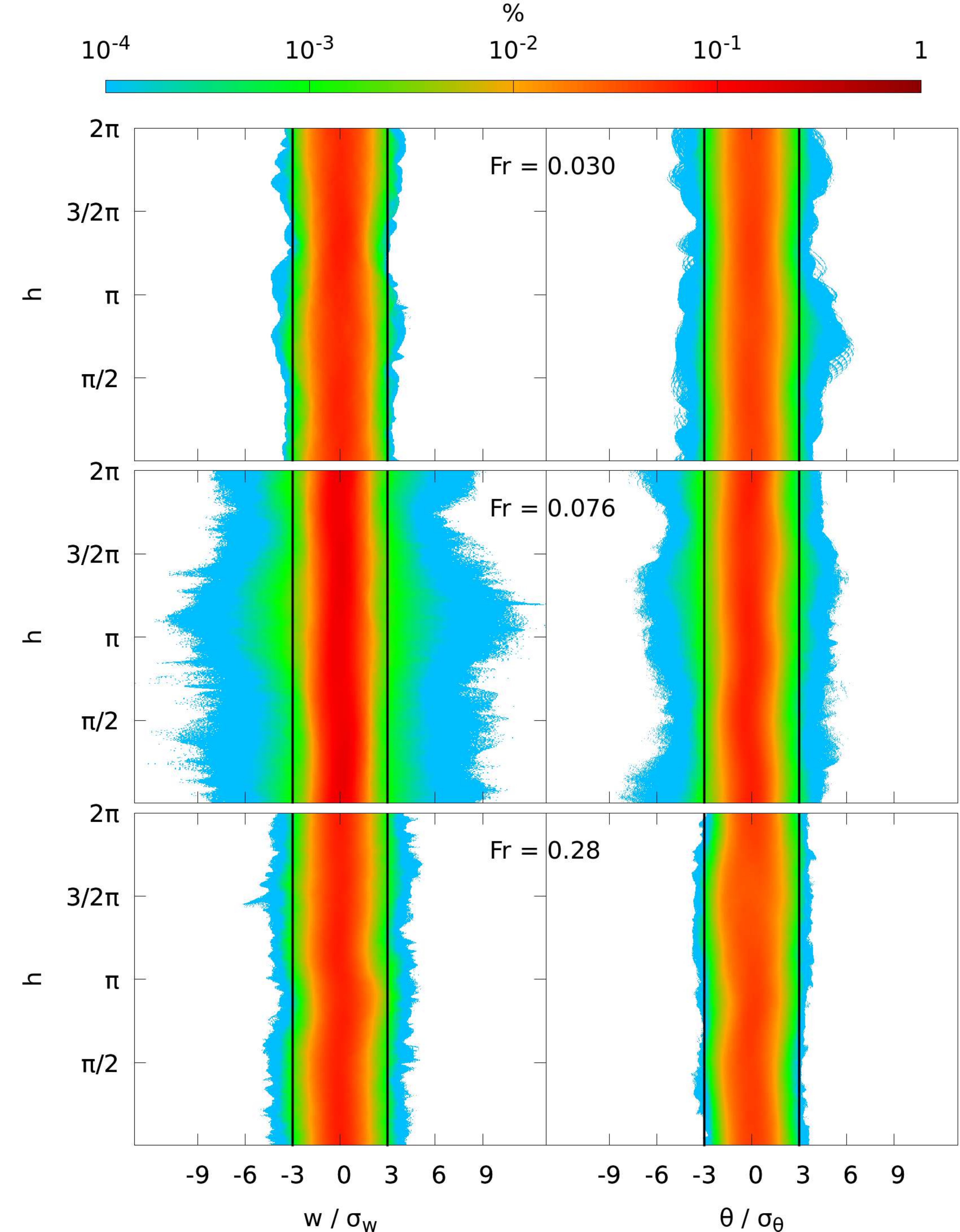}
		\end{center} \vskip-0.2truein   
\caption{\it 2D histograms of $w$ (left) and $\theta$ (right) for three Froude numbers, in bins of their standardized values and of the height $h$ (the $z$-coordinate and direction of gravity). 
The color palette shows the probability that a particular field value has to occur in the plane at altitude $h$. 
Red is for the most common values, light-blue for the rarest (see color bar). The mid panel corresponds to run $9$ which displays extreme events in $w$ and $\theta$. 
Top and bottom panels correspond respectively to runs $3$ 
and  $15$, in which extreme events rarely elop.}
		\label{FIG3}
		\vskip-0.15truein       
	\end{figure}

The PDFs of both $w$ and $\theta$ are shown in Fig.~\ref{FIG2} (left) for run $9$, exhibiting fatter tails than the Gaussian reference. The small asymmetry of the PDFs is a consequence of the limiting sampling, and is expected to disappear if statistics are accumulated over much longer times.
A Gaussian behavior, to the extent that $K_\alpha \approx 3$ ($\alpha=w$, $\theta$), is recovered in the runs with large $Fr$ (weakly stratified) and for $Fr<0.04$ (strongly stratified).
The kurtosis of $u$ and $v$ on the other hand show almost no dependence on $Fr$, with values of 
$K_{u}$ and $K_{v}$ close to $3$ for all runs (see Fig.~\ref{FIG1}). The corresponding PDFs (not shown) are in good agreement with Gaussian or sub-Gaussian distributions.
To characterize the spatial distribution of the intermittent extreme events in $w$ and $\theta$, emerging within the flow with the passage of the time, we computed two-dimensional (2D) histograms of these fields accumulated in time (i.e., using several DNS temporal outputs), displayed in Fig.~\ref{FIG3} for runs $3$, $9$, and $15$. Events are counted as a function of their height and of the standardized values (i.e., normalized to have zero mean and unitary standard deviation) of $w$ and $\theta$, and the counts normalized to get a probability.
The standard deviation $\sigma_\alpha$ ($\alpha=w$, $\theta$) used to normalize each field is computed on the three-dimensional domain at the time of each output.
In this representation, in the absence of extreme events, assuming a Gaussian statistics, 
over 99.7\% of all the points would accumulate in a vertical stripe of the histograms between $\pm 3\alpha/\sigma_\alpha$. However, 
velocity and buoyancy bursts induce the presence of many events with $|\alpha|/\sigma_\alpha > 3$. Histograms for run $9$ are shown in particular in Fig.~\ref{FIG3} (middle) and provide several relevant information: (1) The values of vertical velocity and buoyancy with the highest probability to occur are those between $\pm 3\sigma_\alpha$. However, extreme events (outside that range) 
are rather abundant for both $w$ and $\theta$ for all the runs within the peak of the plot of kurtosis vs.~$Fr$ (Fig.~\ref{FIG1}, top), and this is true in particular for run $9$, for which the probability to observe events larger than $3\sigma_\alpha$ is $1.06\%$ for $\alpha=w$ and $0.85\%$ for $\alpha=\theta$, therefore significantly larger than what expected in the Gaussian case ($0.27 \%$). (2) Extreme events have a higher probability to develop in $w$ than in $\theta$. Moreover, non-zero probabilities are observed in $w$ up to $\approx 13\sigma_w$, whereas in $\theta$ only up to $\approx 9\sigma_\theta$. 
This is in agreement with the fact that peak values of $K_\theta$ are lower than those of $K_w$ (see Fig.~\ref{FIG1}, top). Such difference is probably due to the coupling between $w$ and $\theta$, which is modulated also by other parameters of the system such as $Re$ and $Fr$.
(3) The  pattern described above is also observed in the histograms of the other runs within the range $Fr \in [0.07-0.1]$ (not shown), demonstrating in all cases that extreme events in $w$ and $\theta$ have a probability to develop over time which is rather independent of height, as expected given the flow homogeneity.

Top and bottom panels in Fig.~\ref{FIG3} also show the histograms for two cases outside the peaks of $K_{w,\theta}$ in Fig.~\ref{FIG1}, respectively at lower and higher Froude numbers: $Fr=0.038$ (run $3$), and $Fr=0.28$ (run $15$). The pattern here is significantly different from the one observed for run $9$, with these histograms having almost no points in the region $|\alpha/\sigma_\alpha|>7$
($\alpha=w$, $\theta$), consistent with the low $K_\alpha$ detected and the absence of extreme events in the vertical velocity and buoyancy in these runs.
 \begin{figure*}[t!] 
 \begin{center}
\hspace*{-0.2cm}
 \includegraphics[width=18cm]{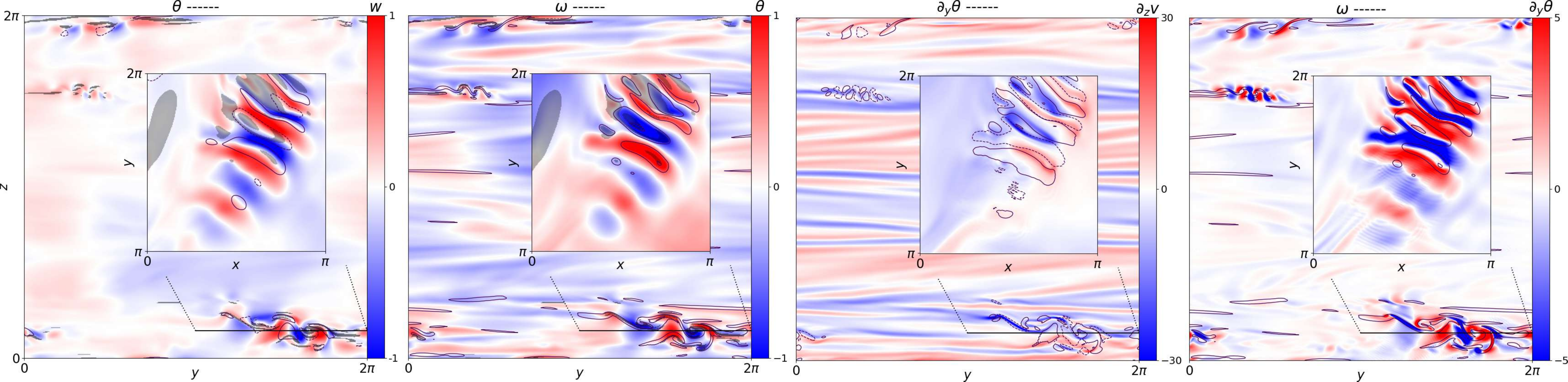}
 \end{center}
 \vskip-0.2truein   
 \caption{\it Vertical 2D slices (with the insets showing a horizontal slice of $1/4$ of the domain at the region of the extreme event) for run $9$. Correlations of $\theta$ and $w$, $\omega = |\boldsymbol{\omega}|$ and $\theta$, $\partial_y \theta$ and $\partial_z v$, and $\omega$ and $\partial_y \theta$ are shown (the first quantity in colors, the second with contours). Contours are at $\pm 3 \sigma_\theta$ for $\theta$, and $\pm 4 \sigma$ for all other fields. Gray shaded areas indicate regions with $\partial_z \theta > N$.}
 \label{FIG4}
 \vskip-0.15truein      
 \end{figure*}
Next we provide evidence of the connection between the large-scale intermittency materializing through the emergence of these bursts of vertical velocity and buoyancy 
\cite{rorai_14,feraco_18}, and the classical internal or small-scale intermittency (evaluated through the kurtosis of the fields and their gradients). From \ref{Tab1} we first note that the kurtosis of all components of the velocity gradient tensor $\partial_i u_j$ (except for ${\partial_z u}$ and ${\partial_z v}$) follow a trend with the Froude number resembling that of the kurtosis of $w$ and $\theta$, with peaks close to $Fr\approx 0.08$. The kurtosis of the diagonal components of the tensor, $K_{\partial_x u}$, $K_{\partial_y v}$, and $K_{\partial_z w}$ are reported in Fig.~\ref{FIG1} (bottom), whereas kurtosis of the off-diagonal terms are in Table \ref{Tab1}, showing how most peak values achieved by $K_{\partial_i u_j}$ are about one order of magnitude larger than the peak values of $K_w$ and $K_\theta$. A similar trend is followed by the kurtosis of the horizontal buoyancy gradient, $K_{\partial_x \theta}$ and $K_{\partial_y \theta}$. Surprisingly $K_{\partial_z \theta}$ (see Fig.\ref{FIG1}, bottom), as the vertical derivatives $K_{\partial_z u}$ and $K_{\partial_z v}$, do not exhibit any definite trend. This can be understood as $\partial_z\theta$ is dynamically bounded and cannot take arbitrarily large values. For $\partial_z\theta \ge N$ buoyancy fluctuations reverse the background stratification, and the flow becomes unstable, developing local convection. Thus, as soon as $\partial_z\theta \approx N$ the flow destabilizes, generating fast and strong large-scale drafts, and $\partial_z\theta$ decreases again \cite{sujovolsky_20}. Indeed, neglecting viscous effects and with $D_t$ the Lagrangian derivative,
\begin{equation}
D_t (\partial_z \theta) = (N-\partial_z \theta) (\partial_z w) - (\partial_x \theta) (\partial_z u) - (\partial_y \theta) (\partial_z v) .
\label{eq:dbdz}
\end{equation}
Strain in horizontal winds ($\partial_z u$ and $\partial_z v$) can change $\partial_z \theta$, but the first term on the r.h.s.~tends to saturate $\partial_z \theta$ as it approaches $N$ \cite{sujovolsky_20}, albeit for this value other gradients grow explosively as the local instability develops.
Given the estimated values of the kurtosis, the PDFs of most gradient tensor elements cannot be Gaussian, as seen in Fig.~\ref{FIG2} (right) for run $9$. This is  a universal feature in turbulent fluids, and a manifestation of small-scale intermittency, with or without stratification. But in this case, the large-scale intermittency and extreme events in $w$ and $\theta$ also enhance the extreme values of the field gradients, thus also enhancing small-scale intermittency. It is also worth noticing that the statistics of the horizontal velocities $u$ and $v$ are always Gaussian (or slightly sub-Gaussian) for all the DNSs in this study, pointing to the fact that the large-scale intermittent behavior of the sole vertical component of the velocity $w$ (and/or of the buoyancy) is able to enhance small-scale intermittency, as seen from the spatial derivatives both of the velocity and of the buoyancy. What is then the origin of the large-scale extreme events, and the link between large-scale intermittent structures and classical small-scale intermittency?

Large-scale extreme events are generated through a buoyancy driven instability, which also connects large- and small-scale dynamics. The visualization in Fig.~\ref{FIG4} ({\it left}) shows localized bursts of $w$, the vertical velocity component, and $\theta$ (\textit{left}), with an essentially periodic structure in the horizontal plane (inset), with a phase difference of $\approx \pi/2$ between the two fields. Fig.~\ref{FIG4} ({\it center left}) also reveals that intense values of $\theta$ coincide with locations where vorticity, $|\bm{\omega}|$, is very strong. These structures can be understood as a result of an instability in the following terms. Vertically sheared horizontal winds (VSHWs) \cite{fitzgerald_18}
are ubiquitous in the flow, as revealed by Fig.~\ref{FIG4} ({\it center right}, with $\partial_z v$ and $\partial_y \theta$), appearing as horizontal streaks with strong values of $\partial_z v$. The resulting shear is prone to Kelvin-Helmholtz instabilities, which creates vertical modulations of the layers, and passively generate horizontal gradients of $\theta$. A nonlinear amplification of the vertical gradients results from the coupling with the velocity gradient, as readily seen from Eq.~(\ref{eq:dbdz}). The induced correlation between, e.g., $\partial_z v$ and $\partial_y \theta$ is clearly seen in the inset of this figure. This mechanism may even lead to values of $\partial_z \theta \ge N$, therefore bringing heavier fluid on top of lighter one, resulting in the formation of intense up- and down-drafts, and to very strong vertical velocity fluctuations (indeed, extreme regions in Fig.~\ref{FIG4} also coincide with regions with $\partial_z \theta \ge N$, as shown by the shaded regions in the first two insets). This process however saturates as the first term on the r.h.s.~of Eq.~(\ref{eq:dbdz}) prevents $\partial_z \theta$ from becoming too large (i.e., the background stratification opposes vertical gradients of $\theta$). Also, from Eqs.~(\ref{bequationV},\ref{bequationT}), a fluid element going up (down) results in an increase (decrease) of $\theta$ following the fluid element, opposing $w$, and thus large updrafts (downdrafts) tend to have a phase shift between $w$ and $\theta$. These motions generate vorticity as 
$D_t \bm{\omega} \approx -N ( \partial_y \theta \hat{\bm x} - \partial_x \theta \hat{\bm y})$ (where viscous effects, forcing and vortex stretching were neglected), resulting in a correlation between $|\bm{\omega}|$ and horizontal gradients of $\theta$, clearly seen in Fig.~\ref{FIG4} ({\it right}, with $|\bm{\omega}|$ and $\partial_y \theta$). In turn, the generation of vorticity feeds the small-scale intermittency, giving the link between large-scale events and internal intermittency. This results in enhanced dissipation, as volumetric dissipation of kinetic energy is proportional to the enstrophy which depends on the spatial average of $\omega^2$. And this also explains the conspicuous absence of vortex filaments in stably stratified turbulence, despite their ubiquity in HIT. In these flows, vortex stretching is \textit{not} the main mechanism producing vorticity (as shown, e.g., by the absence of the Vieillefosse tail \cite{Vieillefosse} in diagrams of invariants of the field gradients \cite{sujovolsky_19, sujovolsky_20}). Instead, generation of vorticity is the result of buoyancy, through a mechanism reminiscent of a large-scale local baroclinic instability.

Thus, large-scale extreme values feed stronger small-scale intermittency through the generation of vorticity. And also the small-scale intermittency can be the precursor of the large-scale extreme values of $w$ and $\theta$, through the billows in the horizontal winds and the amplification of $\partial_z \theta$.

Indeed, this also provides an explanation for the decrease of large-scale intermittency for $Fr<0.04$. As $Fr$ decreases, $N$ increases, and it becomes more difficult for points in the fluid to reach $\partial_z \theta \approx N$. The counterpart of this process in spectral space, and the coupling of these two types of intermittency, could be studied using tools as those used in \cite{alexakis_05, mininni_08} for HIT. Similarities in the behavior of $K_{\partial_x u}$, $K_{\partial_y v}$, and $K_{\partial_z w}$ (and in the kurtosis of the off-diagonal components of $\partial_i u_j$) also suggest that the small scales of the large-scale drafts are in a more mixed state, and in some sense more isotropic (albeit this should not be confused with a recovery of isotropy as in HIT, as, e.g., the vertical velocity has extreme values while horizontal velocities do not, and passive scalars are known in some cases not to return, or to return very slowly, to isotropy), whereas regions outside these patches are less mixed and even more anisotropic. Such more mixed states enforce the connection between these patches and local instabilities.

Finally, this intermittent dynamics at large and small scales is modulated by different regimes explored by the flow, which can be viewed in terms of the relative strength of waves and turbulent motions as measured by $R_B$ (see the labels in Fig.~\ref{FIG1}, indicating the range of $R_B$ for the runs). All simulations with larger values of kurtosis have $R_B \in [10,200]$.
While several theoretical models of small-scale intermittency for the velocity field in HIT have been devised, the intermittency considered here couples multiple scales and combines such intermittency with large-scale bursts. It is thus not clear {\it a priori} how these models perform in geophysical applications or in the presence of waves. It was shown in \cite{rorai_14} and in F18 that a simple one-dimensional model stemming from the original work of Vieillefosse \cite{Vieillefosse}, to which the wave terms were added, behaves remarkably well in reproducing the peak of the kurtosis of the vertical velocity (Fig.~\ref{FIG1}). There are also models that tackle the dynamics of  the passive scalar. For example, in \cite{marro_18}, the dispersion of turbulent plumes in a boundary layer is considered through a model which allows for a prediction of the PDF of the scalar density distribution (see also \cite{bertagni_19}). 
A generalization of such models to the Boussinesq framework was presented in \cite{sujovolsky_19, sujovolsky_20}, and it would be of interest to further extend these models to properly capture the coupling of small-scale and large-scale intermittency in stratified flows. 

\section{Discussion}
\noindent
The aim of the present study was to investigate the large-scale intermittency properties of the velocity and buoyancy fields in forced stably stratified turbulence, and to show its connection with the more classical small-scale (or internal) intermittency. To achieve this goal we exploited a large set of forced DNSs where solutions of the Boussinesq equations were analyzed using statistical tools; in particular, the large-scale intermittency of the velocity and buoyancy fields was evaluated by means of their fourth-order distribution moments (or kurtosis), while the small-scale intermittency was quantified through the kurtosis of field gradients.

Our analysis shows that the buoyancy field $\theta$ (proportional to potential temperature variations) is intermittent at large scales with its kurtosis following the same non-monotonic trend with Froude number as the kurtosis of the vertical component of the velocity $w$.
Thus, well-defined peaks of the kurtosis of both $w$ and $\theta$ appear for $Fr\approx 0.07-0.1$, with values significantly larger than the Gaussian reference of $3$, indicating the emergence of bursts of both quantities in stratified turbulence. 
From the analysis in Fig.~\ref{FIG3} we conclude that the extreme events responsible for this large-scale intermittent behavior in stratified flows can take place at any height and are uniformly distributed along the direction of stratification for long-enough DNS integration times.

Moreover, we showed that small-scale intermittency (i.e., in field gradients) is enhanced in the same range of $Fr$ as large-scale intermittency, with the peaks of their kurtosis occurring for $Fr\approx  0.07-0.1$, and that in individual simulations their extreme values takes place at the same spatial locations as the large-scale patches. We also provided a mechanism for the generation of the large-scale events through the growth of billows in the winds, the amplification of vertical buoyancy gradients through strain, the development of overturning, and the generation of vorticity through baroclinicity which feeds the small scales. This mechanism is consistent with the exact equations for the evolution of the field gradients. Thus, we showed the connection between large-scale extreme events, and small-scale intermittency in the fields $\partial_x u$, $\partial_y v$, $\partial_z w$, the vorticity, and other off-diagonal elements of the velocity gradient, as well as in $\partial_x \theta$ and $\partial_y \theta$. 

Beyond the specific objectives of this study, weather and climate codes in use today require sophisticated modeling through parameterizations of the unresolved small scales. Unraveling the link between small-scale intermittency and large-scale enhancement of vertical velocity and buoyancy is an important element to incorporate in sub-grid models of geophysical flows, and that raise questions for future studies. For example, can these strong vertical shear layers in stratified turbulence (including in the presence of rotation \cite{pouquet_19p}) be modeled adequately with the sub-grid parameterization developed for plane-channel flows, {\it e.g.}, in \cite{leveque_07},  improving on the classical Smagorinsky eddy-viscosity? The fact that the rapid intensification phase of a hurricane is best predicted by following vertical velocity enhancements leads to think that the small-scale and large-scale connection through intermittency is also a factor to be taken into consideration in such models. This might also allow for a more detailed understanding of the interplay between the intermittency observed in some cases in the ocean or in the atmosphere far from the atmospheric or oceanic boundary layers, and the high skewness and kurtosis found in recent oceanic simulations \cite{pearson_18}. For the atmosphere, based on the range of Froude numbers for which strong events develop, typical scales are of the order of $100$ km, while in the ocean the scale is of $10$ km. 
We conclude by stressing that the significance of 
extreme events in the atmosphere goes well beyond the scales considered here~\cite{Rahmstorf:2011}.

\acknowledgments{\it R. Marino acknowledges support from the project ``EVENTFUL" (ANR-20-CE30-0011), funded by the French ``Agence Nationale de la Recherche" - ANR through the program AAPG-2020, and IRP IVMF (CNRS and CONICET). Numerical simulations were done on the cluster ``Newton'' of the High Performance Computing Center of the University of Calabria, supported by EU FP7 2007-13 through the MATERIA Project (PONa3\_00370) and EU Horizon 2020 through the STAR\_2 Project (PON R\&I 2014-20, PIR01\_00008). AP is thankful to LASP and Bob Ergun.}


\begin{thebibliography}{999}

\bibitem{frisch_80} FRISCH U., Fully developed turbulence and intermittency. {\em Annals New York Acad. Sci.} {\bf 1980}, {\em 357}, 359. 
\bibitem{fritts_13} FRITTS D.C. and WANG L., Gravity wave-fine structure interactions. Part I: Influences of fine structure form and orientation on flow evolution and instability. {\em J. Geophys. Res.} {\bf 2013}, {\em 70}, 3735.
\bibitem{vanharen_15} VAN HAREN H., CIMATORIBUS A. and GOSTIAUX L. , Where large deep‐ocean waves break. Geophys. {\em Res. Lett.}, {\bf 2015}, {\em 42}, 2351– 2357.
\bibitem{pearson_18} PEARSON B. and  FOX-KEMPER B., Log-normal turbulence dissipation in global ocean models. {\em Phys. Rev. Lett.}  {\bf 2018}, {\em 120}, 094501. 
\bibitem{kolmogorov_62} KOLMOGOROV A.N., A refinement of previous hypotheses concerning the local structure of turbulence in a viscous incompressible fluid at high Reynolds number. {\em J. Fluid Mech.} {\bf 1962}, {\em 13}, 82.
\bibitem{she_94} SHE Z-S. and L\'EV\^EQUE E., Universal scaling laws in fully developed turbulence.  {\em Phys. Rev. Lett.} {\bf 1994}, {\em 72}, 336. 
\bibitem{toschi_00} TOSCHI F., L\'EV\^EQUE E. AND RUIZ-CHAVARRIA G., Shear Effects in Nonhomogeneous Turbulence. {\em  Phys. Rev. Lett.} {\bf 2000}, {\em 85}, 1436.

\bibitem{mahrt_89} MAHRT L., Intermittency of atmospheric turbulence. {\em J. Atmosph. Sci.} {\bf 1989}, {\em 46}, 79.

\bibitem{Bramberger_2020} BRAMBERGER M., D{\"O}RNBRACK A., WILMS H., EWALD F., AND SHARMAN, R., Mountain-Wave Turbulence Encounter of the Research Aircraft HALO above Iceland. {\em Journal of Applied Meteorology and Climatology} {\bf 2020}, {\bf 59}, {\em 567}.
\bibitem{lyu_2018} LYU R., HU F., LIU L., XU J. and CHENG X., High-order statistics of temperature fluctuations in an unstable atmospheric surface layer over grassland. {\em Advances in Atmospheric Sciences}  {\bf 2018}, {\em 35}, 1265.

\bibitem{chau_17} CHAU J.L., STOBER G., HALL C.M., TSUTSUMI M., LASKAR F.I. and HOFFMANN P., Polar mesospheric horizontal divergence and relative vorticity measurements using multiple specular meteor radars. {\em Radio Science} {\bf 2017}, {\em 52}, 811.

\bibitem{dasaro_07} D'ASARO E., LIEN R. and HENYEY F., High-Frequency internal waves on the Oregon continental shelf. {\em J. of Phys. Oceanogr.} {\bf 2011}, {\em 332}, 318.
\bibitem{rorai_14} RORAI C., MININNI P.D. and POUQUET A., Turbulence comes in bursts in stably stratified flows. {\em Phys. Rev. E} {\bf 2014}, {\em 89}, 043002.
\bibitem{feraco_18} FERACO F., MARINO R., PUMIR A., PRIMAVERA L., MININNI P.D., POUQUET A. and ROSENBERG D., Vertical drafts and mixing in stratified turbulence: Sharp transition with
Froude number. {\em Eur. Phys. Lett.} {\bf 2018}, {\em 123}, 44002.
\bibitem{buaria} 
D. BUARIA, A. PUMIR, F. FERACO, R. MARINO, A. POUQUET, D. ROSENBERG, and L. PRIMAVERA,  Single-particle Lagrangian statistics from direct numerical simulations of rotating-stratified turbulence. {\em Phys. Rev. Fluids} {\bf 2020}, {\em 5}, 064801.
\bibitem{capet_08} CAPET X., MCWILLIAMS J.C., MOLEMAKER M.J. and SHCHEPETKIN A.F., Mesoscale to submesoscale transition in the California current system. Part I: flow structure, eddy flux, and observational tests. {\em J. Phys. Oceanogr.} {\bf 2008}, {\em 38}, 29.

\bibitem{fontana_20} FONTANA M., BRUNO P.O., MININNI P.D. and DMITRUK P., Fourier continuation method for incompressible fluids with boundaries. {\em Computer Physics Communications} {\bf 2020}, {\em 256}, 107482.

\bibitem{rosenberg_20}
ROSENBERG D., MININNI P.D., REDDY R. and POUQUET, A., GPU Parallelization of a Hybrid Pseudospectral Geophysical Turbulence Framework Using CUDA. {\em Atmosphere} {\bf 2020}, {\em 11}, 00178.

\bibitem{sujovolsky_19} SUJOVOLSKY N.E. and MININNI P.D., Invariant manifolds in stratified turbulence. {\em Phys. Rev. Fluids} {\bf 2019} {\em 4},052402.

\bibitem{marino_14} MARINO R., MININNI P.D., ROSENBERG D., POUQUET A., Large-scale anisotropy in stably stratified rotating flows. {\em Phys. Rev. E}, {\bf 2014}, {\bf 90}, 023018.
\bibitem{pouquet_19p} 
POUQUET A., ROSENBERG D. and MARINO, R.,  Linking dissipation, anisotropy and intermittency in rotating stratified turbulence. {\em Phys. Fluids} {\bf 2019}, {\em 31}, 105116.

\bibitem{sujovolsky_20}
SUJOVOLSKY, N.E. and MININNI, P.D., From waves to convection and back again: The phase space of stably stratified turbulence, {\em Phys. Rev. Fluids}, {\bf 2020}, {\bf 5}, 064802.
\bibitem{fitzgerald_18} FITZGERALD J. G. and FARRELL B. F., Vertically Sheared Horizontal Flow-Forming Instability in Stratified Turbulence: Analytical Linear Stability Analysis of Statistical State Dynamics Equilibria, {\em Journal of the Atmospheric Sciences}, {\bf 2018}, {\em 75(12)}, 4201-4227.
\bibitem{alexakis_05} ALEXAKIS A., MININNI P.D. and POUQUET A., Imprint of large-scale flows on Navier-Stokes turbulence. {\em Phys. Rev. Lett.}  {\bf 2005}, {\em 95}, 264503.
\bibitem{mininni_08} MININNI P.D., ALEXAKIS A. and POUQUET A., Nonlocal interactions in hydrodynamic turbulence at high {R}eynolds numbers: {T}he slow emergence of scaling laws. {\em Phys. Rev. E}  {\bf 2008}, {\em 77}, 036306.
\bibitem{marro_18} MARRO M., SALIZZONI P., SOULHAC, L. and CASSIANI, M.,
Dispersion of a Passive Scalar Fluctuating Plume in a Turbulent Boundary Layer. Part {III}: Stochastic Modelling.
{\em Bound. Layer Met.}  {\bf 2018},  {\em 167}, 349.
\bibitem{Vieillefosse} VIEILLEFOSSE P., {\em Physica} A {\bf 125} (1984) 150.
\bibitem{bertagni_19} BERTAGNI, M.B., MARRO, M., SALIZZONI, P. and CAMPOREALE, C.,
Solution for the statistical moments of scalar turbulence. {\em Phys. Rev. Fluids}  {\bf 2019},  {\em 4}, 124701.
\bibitem{leveque_07} L\'EV\^EQUE, E., TOSCHI, F., SHAO, L. and BERTOGLIO, J.-P.,
Shear-improved Smagorinsky model for large-eddy simulation of wall-bounded turbulent flows. 
{\em J. Fluid Mech.} {\bf 2007}, {\em 570}, 491.
\bibitem{Rahmstorf:2011} RAHMSTORF, S. and COUMOU, D.,
Increse of extreme events in a warming world.
{\em PNAS} {\bf 2011}, {\em 108}, 17905-17906.

\end{thebibliography}
\end{document}